\documentstyle[11pt]{article}     

\expandafter\ifx\csname amssym.def\endcsname\relax \else\endinput\fi
\expandafter\edef\csname amssym.def\endcsname{%
       \catcode`\noexpand\@=\the\catcode`\@\space}
\catcode`\@=11
\def\undefine#1{\let#1\undefined}
\def\newsymbol#1#2#3#4#5{\let\next@\relax
 \ifnum#2=\@ne\let\next@\msafam@\else
 \ifnum#2=\tw@\let\next@\msbfam@\fi\fi
 \mathchardef#1="#3\next@#4#5}
\def\mathhexbox@#1#2#3{\relax
 \ifmmode\mathpalette{}{\m@th\mathchar"#1#2#3}%
 \else\leavevmode\hbox{$\m@th\mathchar"#1#2#3$}\fi}
\def\hexnumber@#1{\ifcase#1 0\or 1\or 2\or 3\or 4\or 5\or 6\or 7\or 8\or
 9\or A\or B\or C\or D\or E\or F\fi}

\font\tenmsa=msam10
\font\sevenmsa=msam7
\font\fivemsa=msam5
\newfam\msafam
\textfont\msafam=\tenmsa
\scriptfont\msafam=\sevenmsa
\scriptscriptfont\msafam=\fivemsa
\edef\msafam@{\hexnumber@\msafam}
\mathchardef\dabar@"0\msafam@39
\def\dashrightarrow{\mathrel{\dabar@\dabar@\mathchar"0\msafam@4B}}
\def\dashleftarrow{\mathrel{\mathchar"0\msafam@4C\dabar@\dabar@}}

\def\ulcorner{\delimiter"4\msafam@70\msafam@70 }
\def\urcorner{\delimiter"5\msafam@71\msafam@71 }
\def\llcorner{\delimiter"4\msafam@78\msafam@78 }
\def\lrcorner{\delimiter"5\msafam@79\msafam@79 }
\def\yen{{\mathhexbox@\msafam@55}}
\def\checkmark{{\mathhexbox@\msafam@58}}
\def\circledR{{\mathhexbox@\msafam@72}}
\def\maltese{{\mathhexbox@\msafam@7A}}
\def\circledS{{\mathhexbox@\msafam@73}}

\font\tenmsb=msbm10
\font\sevenmsb=msbm7
\font\fivemsb=msbm5
\newfam\msbfam
\textfont\msbfam=\tenmsb
\scriptfont\msbfam=\sevenmsb
\scriptscriptfont\msbfam=\fivemsb
\edef\msbfam@{\hexnumber@\msbfam}
\def\Bbb#1{{\fam\msbfam\relax#1}}
\def\widehat#1{\setbox\z@\hbox{$\m@th#1$}%
 \ifdim\wd\z@>\tw@ em\mathaccent"0\msbfam@5B{#1}%
 \else\mathaccent"0362{#1}\fi}
\def\widetilde#1{\setbox\z@\hbox{$\m@th#1$}%
 \ifdim\wd\z@>\tw@ em\mathaccent"0\msbfam@5D{#1}%
 \else\mathaccent"0365{#1}\fi}
\font\teneufm=eufm10
\font\seveneufm=eufm7
\font\fiveeufm=eufm5
\newfam\eufmfam
\textfont\eufmfam=\teneufm
\scriptfont\eufmfam=\seveneufm
\scriptscriptfont\eufmfam=\fiveeufm
\def\frak#1{{\fam\eufmfam\relax#1}}

\csname amssym.def\endcsname

\makeatletter
\def\section{\@startsection {section}{1}{\z@}{-3.5ex plus -1ex minus 
 -.2ex}{2.3ex plus .2ex}{\large\sc}}
\def\subsection{\@startsection{subsection}{2}{\z@}{-3.25ex plus -1ex minus 
 -.2ex}{1.5ex plus .2ex}{\normalsize\sc}}
\makeatother

\makeatletter
\@addtoreset{equation}{section}                        

\makeatother

\newcommand{\nc}{\newcommand}
\newcommand{\rnc}{\renewcommand}

\nc{\be}{\begin{equation}}
\nc{\ee}{\end{equation}}
\nc{\bea}{\begin{eqnarray}}
\nc{\eea}{\end{eqnarray}}

\nc{\trac}[2]{{\textstyle\frac{#1}{#2}}}

\nc{\ex}[1]{\mbox{e}^{\,\textstyle#1}}

\nc{\CC}{\Bbb{C}}
\nc{\HH}{\Bbb{H}}
\nc{\PP}{\Bbb{P}}
\nc{\RR}{\Bbb{R}}
\nc{\ZZ}{\Bbb{Z}}
\nc{\II}{\Bbb{I}}
\nc{\EE}{\Bbb{E}}
\nc{\SS}{\Bbb{S}}

\def\slash#1{\setbox0=\hbox{$#1$}#1\hskip-\wd0\hbox to\wd0{\hss\sl/\/\hss}}
 
\rnc{\a}{\alpha}
\nc{\al}{\a^{l}}
\rnc{\d}{\delta}
\nc{\ga}{\gamma}
\nc{\la}{\lambda}
\nc{\lal}{\la_{l}}
\nc{\f}{\phi}
\nc{\fb}{\bar{\phi}}
\nc{\p}{\psi}

\nc{\e}{\eta}
\nc{\eb}{\bar{\eta}}
\rnc{\c}{\chi}
\nc{\eps}{\epsilon}
\rnc{\t}{\theta}
\nc{\tb}{\bar{\theta}}
\nc{\om}{\omega}
\nc{\Om}{\Omega}

\rnc{\P}{\Psi}
\nc{\pl}{\P_{L}}
\nc{\pdr}{\P^{\dag}_{R}}
\nc{\G}{\Gamma}

\nc{\sig}{\sigma}
\nc{\sk}{\sigma_{k}}
\nc{\sa}{\sigma_{a}}
\nc{\Bb}{\bar{B}}

\nc{\symx}{\circledS}

\nc{\Q}{\bar{Q}}
\nc{\M}{{\cal M}}                          
\nc{\C}{{\cal A}/{\cal G}}                
\nc{\A}[1]{{\cal A}^{#1}/{\cal G}^{#1}}  
\nc{\RC}{{\cal R}_{\C}}                 
\nc{\RM}{{\cal R}_{\M}}                
\nc{\RX}{{\cal R}_{X}}
\nc{\RY}{{\cal R}_{Y}}

\nc{\ad}{\mathop{\mbox{ad}}\nolimits}
\nc{\tr}{\mathop{\mbox{tr}}\nolimits}
\nc{\Tr}{\mathop{\mbox{Tr}}\nolimits}
\nc{\Det}{\mathop{\mbox{Det}}\nolimits}
\rnc{\det}{\mathop{\mbox{det}}\nolimits}
\nc{\rk}{\mathop{\mbox{rk}}\nolimits}
\nc{\diag}{\mbox{diag}}
\nc{\ra}{\rightarrow}
\nc{\Ra}{\Rightarrow}
\nc{\LRa}{\Leftrightarrow}
\nc{\lra}{\leftrightarrow}
\nc{\ot}{\otimes}
\rnc{\ss}{\subset}
\nc{\nul}{\noindent\underline}
\nc{\non}{\nonumber\\}
\rnc{\S}{\Sigma}
\nc{\tp}{2\pi i}
\nc{\del}{\partial}
\nc{\dbar}{\bar{\del}}
\nc{\dx}{\dot{x}}
\rnc{\lg}{\frak{g}}

\nc{\zb}{\bar{z}}
 
\nc{\mat}[4]{\left(\begin{array}{cc}#1&#2\\#3&#4\end{array}\right)}
 
\nc{\r}[1]{\mathbf{#1}}
\nc{\rb}[1]{\overline{\mathbf{#1}}}

\nc{\gi}{\gamma_{i}}
\nc{\gj}{\gamma_{j}}

\nc{\subs}[1]{{\vspace*{0.5cm}}%
{\noindent\underline{\small\sc #1}}{\addcontentsline{toc}{subsubsection}{#1}}%
{\vspace*{0.3cm}}}

\nc{\chap}[1]{{\clearpage}%
\begin{center}%
{\noindent\underline{\large\sc #1}}{\addcontentsline{toc}{section}{#1}}%
\end{center}%
{\vspace*{0.3cm}}}

\begin{document}
\global\parskip=4pt

\begin{titlepage}
\begin{flushright}
\rightline{hep-th/9706225}
\end{flushright}
\vspace*{0.5in}
\begin{center}
{\LARGE\sc Euclidean SYM Theories By Time Reduction \\[4mm]
And Special Holonomy Manifolds}\\
\vskip .3in
\begin{tabular}{cc}
{\sc Matthias Blau}\footnotemark
&
{\sc George Thompson}\footnotemark 
\\[.1in]
LPTHE-{\sc enslapp}\footnotemark
&
ICTP\\
ENS-Lyon,
46 All\'ee d'Italie
&
P.O. Box 586\\
F-69364 Lyon CEDEX 07, France
&
34014 Trieste, Italy\\
\end{tabular}
\end{center}
\addtocounter{footnote}{-2}%
\footnotetext{e-mail: mblau@enslapp.ens-lyon.fr}
\addtocounter{footnote}{1}%
\footnotetext{e-mail: thompson@ictp.trieste.it}
\addtocounter{footnote}{1}%
\footnotetext{URA 14-36 du CNRS, associ\'ee \`a l'E.N.S. de Lyon,
et \`a l'Universit\'e de Savoie}
\vskip .50in
\begin{abstract}
\noindent Euclidean supersymmetric 
theories are obtained from Minkowskian theories
by performing a reduction in the time direction. This procedure 
elucidates certain mysterious features of Zumino's $N\!=\!2$ model in four 
dimensions, provides manifestly hermitian Euclidean counterparts of all 
non-mimimal SYM theories, and is also applicable to supergravity theories.
We reanalyse the twists of the 4d $N\!=\!2$ and 
$N\!=\!4$ models from this point of view. Other applications include 
SYM theories on special holonomy manifolds. In particular, we construct
a twisted SYM theory on K\"ahler 3-folds and clarify the structure of
SYM theory on hyper-K\"ahler 4-folds.
\end{abstract}
\end{titlepage}

\begin{small}
\tableofcontents
\end{small}

\setcounter{footnote}{0}

\section{Introduction}

This paper is concerned with a number of related issues. In one way or
another they are all tied to the problem of finding Euclidean
supersymmetric Yang-Mills (SYM) theories. The way we will produce these
theories is by dimensional reduction of Minkowskian theories along some
directions one of which is time. This construction leads to manifestly
hermitian Euclidean SYM actions and thus to viable Euclidean
counterparts of all non-minimal Minkowskian SYM theories. It reproduces
the known Euclidean SYM actions, e.g.\ Zumino's $N\!=\!2$ supersymmertic
instanton theory in $d\!=\!4$ \cite{zumino}, and explains in a natural way
the features one has come to expect of such theories, such as
non-compact R-symmetry groups and one scalar field whose kinetic term
has the wrong sign (for a recent review of these matters see
\cite{vanN}).

One place where Euclidean SYM theories play an important role is in the
context of topological field theories. We reanalyse the known twists
of the $d\!=\!4$ $N\!=\!2$ \cite{ewdon} and 
$N\!=\!4$ \cite{yamron,vw,marcus,bsv,btn2,ll}
theories by twisting directly the Euclidean SYM actions via the 
non-compact R-symmetry groups and show that the reality properties of
the fields (which are usually introduced by hand) and of the action of the
twisted theory are a consequence of the hermiticity of the underlying 
Euclidean SYM action.

Euclidean SYM actions also appear naturally in the context of 
D-brane instantons in string theory\footnote{In this context, 
the appearance of a scalar field with the `wrong' sign alluded
to above is natural as it should correspond to the time-like collective 
coordinate of the instanton, in the spirit of \cite{ewpb}.} 
\cite{bsv,btn2}, and for this
and related reasons there has been some interest recently in 
(twisted) Euclidean SYM theories on $d>4$ manifolds with special
holonomy groups \cite{bks,aos,aos2}. These provide field theory realizations
of the generalized self-duality conditions studied in \cite{cdfn,ward}
and hence a topological field theory counterpart of the mathematical
research programme outlined in \cite{dt}. Here we survey these
constructions and attempt to clarify the relation between twisted theories
and theories on manifolds possessing covariantly constant spinors.
We point out the existence of a new twisted field theory on K\"ahler
three-folds (modelling the Donaldson-Uhlenbeck-Yau equations), and
describe the theory (and generalized self-duality condition) one 
obtains on hyper-K\"ahler eight-manifolds.

\section{Euclidean SYM Theories from Time Reduction}

Many years ago Zumino wrote down a Euclidean version of the $N\!=\!2$ SYM
theory in four dimensions \cite{zumino}. 
There are two rather striking features of
this model. The first is that one of the two scalars (a true scalar) has
a kinetic term with the wrong sign, whereas the other (the
pseudo-scalar) has the standard sign. With the Minkowskian metric no
such `degenerate' behaviour is present. The second surprise is that in
the Minkowskian theory there is a chiral symmetry that is compact while the
Lorentz group $SO(3,1)$ is not whereas in the Euclidean theory the
situation is reversed, the chiral symmetry group is non-compact while
the Lorentz group $SO(4)$ is compact. These differences can be accomodated by
starting with a single theory in $6$ dimensions. The
observation is quite simple and also has a straightforward generalization
to other Euclidean SYM theories.

\subsection{Zumino's Model from $N\!=\!1$ $d\!=\!(5+1)$ SYM}

Let us first describe the procedure in words.
In $d\!=\!6$ one starts with the usual Minkowskian $N\!=\!1$ SYM
theory. The field content of this  theory consists of one six-vector 
(the gauge field) and one Weyl fermion in the $\r4$ of $SO(5,1)$. Apart
from the gauge group and the $N\!=\!1$ supersymmetry the model enjoys an
$SU(2)_{{\cal R}}$ symmetry.

Now suppose that we dimensionally reduce in the standard way along two space
directions. In this way we will obtain an $N\!=\!2$ theory with
Min\-kow\-ski\-an signature in four dimensions. The Lorentz group decomposes
as $SO(5,1) \to SO(3,1) \times SO(2)$, where the $SO(2)$ is an
internal symmetry group from the four dimensional stance. It is this
$SO(2) \sim U(1)$ that plays the role of the compact chiral symmetry
alluded to above. The two scalars that one gets from the six vector
transform under the $SO(2)$ while the four-vector remains inert. The
two scalars are naturally grouped into one complex scalar. In this way
we have recovered all the features of
the four-dimensional theory from the six-dimensional one. 

Now let us 
perform a variant on this theme. We dimensionally reduce in one
spatial direction and in the time direction. The resulting theory is a
four dimensional Euclidean theory. There are two scalars here as well
but one has the wrong sign (the one that is the time component of the
gauge field). The Lorentz group now decomposes as $SO(5,1) \to SO(4)
\times SO(1,1)$. The $SO(1,1)$ is now the non-compact chiral symmetry
group. Once more the general features of Zumino's theory are reproduced. 

One can perform the reductions explicitly and since we will need the
fermionic action anyway we present
the, perhaps unfamiliar, time reduction here. We begin in 6
dimensions. The spinor part of the $N\!=\!1$ theory is
\be
\int d^{6}x \, \overline{\psi}_{6} \slash{D} \psi_{6} \label{spin6}
\ee
where $\psi$ is a Weyl spinor. A basis for the $\ga$-matrices can be
found in the Appendix. We `compactify' (i.e.\ dimensinally reduce along)
the time-direction $t=x_{0}$.
Doing so one should pick up a Euclidean theory in 5 dimensions. In
particular the spinor content should be that of a four dimensional
Euclidean Dirac spinor, with the usual action. At first, however, the action
(\ref{spin6}), becomes
\be
\int d^{5}x \, \psi^{*T}_{6} \Gamma^{0}\left[ \Gamma^{m}D_{m} +
\Gamma^{0}A_{0} \right] \psi_{6}  \label{spin5}
\ee
which has a seemingly spurious $\Gamma_{0}$ in it. But the spinor is
chiral so we can write it as
\be
\psi_{6} = \left( \begin{array}{c}
\psi_{E} \\
0
\end{array} \right)
\ee
and the action  (\ref{spin5}) becomes
\be
\int d^{5}x \, \psi^{*T}_{E} \left[ \gamma^{m}D_{m} - A_{0} \right]
\psi_{E}   \label{spin5a}
\ee
Upon further reduction along $x_{5}$ to 4 dimensions one obtains
\be
\int d^{4}x \, \psi^{*T}_{E} \left[ \gamma^{\mu}D_{\mu} + \ga_{5}A_{5}
- A_{0} \right]  \psi_{E} .  \label{spin4}
\ee
Note that the field
$A_{0}$ is the one that comes in with the wrong sign in the kinetic
term, Zumino calls it $A$. The coupling we have found in
(\ref{spin5a}) between the fermions and the (pseudo-) scalars $A_{5}$ and
$A_{0}$ agrees with that given by Zumino \cite{zumino}. One can easily
check that the other parts of the action also agree.

We would now like to compare our dimensional reduction point of view
with some recent work on Wick rotations \cite{vanN}. These authors
show that a consistent manner for obtaining Zumino's Euclidean theory
from the Minkowski theory is to perform a Wick rotation. This
rotation, it is claimed, should have the interpretation of being a complex
Lorentz transformation in a would-be (0,4)-plane (with the above 
notation). The authors, though 
they give more arguments as to why their prescription should be thought of
as a five dimensional Lorentz transformation, admit that there are
some `loose ends' which need to be fixed. From our higher-dimensional
point of view we see that they are indeed morally correct. 

One starts in 5-dimensional Minkowski space (and to conform with the
notation in \cite{vanN} we assume a space
reduction of the 6-dimensional theory along $x_{4}$). The coordinates are
thus $(t=x_{0},x_{i},x_{5})$ with $i = 1, 2,3$. One may pass to the four
dimensional Minkowski space by `fixing' $x_{5}=0$. All the fields then
have the functional dependence $\psi (t, x_{i}, 0)$. From the five
dimensional perspective one can now perform a complex Lorentz boost in the
$(t,x_{5})$ plane mapping $(t,x_{5})$ to $(ix_{5}, it)$. For the fermions 
one finds
\be
\psi(t,x_{i},0) \to \ex{\frac{\pi}{2}i\Gamma_{0}\Gamma_{5}}\psi(0,x_{i},it)
\ee
which is precisely the observation in \cite{vanN}. As regards the
gauge fields, from the point of view of four-dimensional Minkowski space, 
$A_{5}$ is a scalar. However, the rotation
defined above exchanges $A_{0}$ with $A_{5}$ and, because of the $i$
factor, $A_{5}$ combines with $A_{i}$ to form a Euclidean four-vector.
$A_{0}$ is now a scalar with the wrong sign for the kinetic
term, exactly as in the above derivation of Zumino's model. 
Notice also that the pseudo-scalar, which was the space component
$A_{4}$ of the gauge field, does not transform.

Of course the point of the paper \cite{vanN} is to give a formulation
of Euclidean supersymmetry even when there is no extended
supersymmetry in the Min\-kow\-ski\-an theory, and this certainly goes
beyond what we are able to do here.

\subsection{R-Symmetries, Chiral Symmetries, and Twisting}

The $SU(2)_{{\cal R}}$ symmetry in $d\!=\!6$ is made up of a `charge
conjugation' symmetry 
\be
\d \psi_{6} = {\cal C}\psi^{*}_{6}
\ee
(see the Appendix) as well as a normal phase transformation
\be
\d\psi_{6} = i \psi_{6}
\ee
and their commutator. 
Thus the single Weyl fermion (together with its complex conjugate)
transforms as a doublet of $SU(2)_{{\cal R}}$. 
It can thus be thought of as a $(\r4,\r2)$ of $SO(5,1)\times SU(2)_{{\cal
R}}$ satisfying a symplectic Majorana-Weyl condition \cite{kt}, which
is possible
since both representations are pseudo-real. This gives the four on-shell
degrees of freedom required to match those of the gauge field.

This symmetry survives the dimensional reduction
regardless of the signature that one chooses for the reduced space.
Upon reduction to four Euclidean dimensions, the $\r4$ of $SO(5,1)$ 
branches to 
the $(\r2,\r1)\oplus(\r1,\r2)$ of $SO(4)\sim SU(2)_{L}\times SU(2)_{R}$.
If we write
\be
\Psi_{L} = \left( \begin{array}{c}
              \psi_{L} \\
\sigma_{2} \psi_{L}^{*}
\end{array} \right), \;\;\; 
\Psi_{R} = \left( \begin{array}{c}
              \psi_{R} \\
\sigma_{2} \psi_{R}^{*}
\end{array} \right)
\ee
then the $SU(2)_{{\cal R}}$ symmetry takes the simple form
\be
\d_{i} \Psi_{L.R} = (\sigma_{i}\otimes\II)\Psi_{L,R} .
\ee
Note that while there are no Majorana spinors in four Euclidean
dimensions, this being the obvious obstacle to formulating
Euclidean $d\!=\!4$ SYM theories, 
the four-dimensional spinors one obtains in this way satisfy a 
symplectic Majorana condition: they are symplectic
Majorana-Weyl with respect to this $SU(2)_{{\cal R}}$. 

Note also that, because of the $\sigma_{2}$ in the definition of
the $\psi_{L.R}$, they tranform essentially the same way under Lorentz
transformations as under the R-symmetry. Thus the twisting is
straightforward directly within the Euclidean setting and one
obtains the familiar Grassmann odd field content
\bea
(2,1,2)\oplus(1,2,2) &\ra& (2,2)\oplus (3,1) \oplus (1,1)\non
                     &\simeq& (\psi,\chi^{+},\eta)
\eea
and scalar supercharge of Donaldson theory \cite{ewdon}.

The ghost number symmetry arises on dimensional reduction. 
The precise form of the symmetry group depends on whether one does a
space-space or space-time reduction. With a space-space reduction one
picks up an $SO(2)$ symmetry which translates into a $U(1)$ symmetry
if we group the two scalar fields into one complex field $\f$.
The action is
\be
\int d^{4}x \f \Delta \bar{\f} \label{scalaraction}
\ee
which is clearly $U(1)$ invariant. On the other hand, performing a
space-time reduction we would also get an action of the form
(\ref{scalaraction}) except that $\f$ and $\bar{\f}$ appearing there are
two independent {\em real} fields. The ghost symmetry is now $SO(1,1)$
which is simply an invariance of the action under scaling $\f \to \la
\f$ and $\bar{\f} \to \la^{-1} \bar{\f}$ for $\la \in {\Bbb R}^{*}$.
For the fermions, in the reduction to Euclidean space, one finds that
the $SO(1,1)$ transformations act as
\be
\d \psi_{6} = \la \Sigma_{50}\psi_{6} = \la \left( \begin{array}{cc}
-\ga_{5} & 0 \\
0 & \ga_{5} 
\end{array} \right)\psi_{6}
\ee
(this being the Lorentz generator in the $(50)$ direction) 
once more in agreement with \cite{zumino}. 

Within the functional integral approach to
Donaldson theory \cite{ewdon}, the two scalar fields $\f$ and
$\bar{\f}$ are nevertheless usually treated as complex conjugates
(at the expense of hermiticity of the action),
and it is then legitimate to regard the ghost number symmetry 
as being $U(1)$ instead of $SO(1,1)$, as is usually done.

Finally, note that on four-manifolds admitting covariantly constant
spinors (for $d\!=\!4$ this singles out K3 and $T^{4}$), the untwisted 
Euclidean SYM theory possesses a supersymmetry precisely of the 
same type (scalar supercharges) as the twisted theories above. In
that case (we would say that) no twist is required. In this sense
we differ semantically from the presentation in \cite{bks}. We will
have more to say about such theories (in particular in $d>4$) in
section 3. Other possibilities (partially twisted topological theories)
can arise when the four-manifold $M_{4}$
is not irreducible but rather a product $M_{4}=M_{3}\times S^{1}$
or $M_{4}=M_{2}\times M_{2}'$ (see e.g.\ \cite{bjsv}).

\subsection{General Properties}

It should be clear that the method outlined above to obtain a 
hermitian Euclidean SYM theory in $d\!=\!4$, time 
reduction of the Minkowskian $N\!=\!1$ theory in $d\!=\!6$ to $d\!=\!4$,
generalizes in a straightforward to other SYM theories: starting
with a SYM theory in $(d+p+1)$-dimensional Minkowski space,
one can obtain a hermitian Euclidean SYM action in $d$ Euclidean
dimensions by simply reducing along $p$ space-like and the one time-like
direction. This is the Euclidean counterpart of the Minkowskian
SYM theory obtained in the standard way by dimensional reduction
along $(p+1)$ space-like directions. The Euclidean theory obtained
in this way will have a non-compact internal R-symmetry group
$SO(p,1)$ and, as before, there will be fields (among them the
time-component of the gauge field) having a kinetic term with the
wrong sign. 

Although in the above we have described the procedure for SYM theories.
it clearly works equally well for other supersymmetric theories such as
$N\!=\!1$ supergravity in $d\!=\!(10+1)$ which can be reduced to a Euclidean 
supergravity theory in $d\!=\!10$, 
a Euclidean analogue of type IIA supergravity.

For concreteness, however, 
and because it is of the most interest in the current
endeavour to understand higher-dimensional SYM theories and the 
world-volume actions of D-branes, let us consider the $N\!=\!1$ SYM
theory in $d\!=\!9+1$ dimensions. Its field content consists of a 
gauge field and a Majorana-Weyl fermion in the $\r{16}$ of $SO(9,1)$.
As a consequence there is no R-symmetry group in ten dimensions.

Standard dimensional reduction leads to Minkowskian SYM theories with
16 supercharges in $d< 10$ with compact R-symmetry group $SO(10-d)$
arising form the branching 
\be
SO(9,1)\ra SO(d-1,1) \times SO(10-d)\;\;.
\ee
The most prominent member of this hierarchy of theories is the scale
invariant and presumably S-dual $N\!=\!4$ theory in 
$d\!=\!3+1$ with R-symmetry group $SO(6)\sim SU(4)$.

By following the above procedure,
(reduction from $(9+1)$ to $d$ Euclidean dimensions),
one obtains Euclidean counterparts of these theories for all $d<10$.
The relevant branching this time is
\be
SO(9,1)\ra SO(d) \times SO(9-d,1)\;\;,
\ee
exhibiting 
$SO(9-d,1)$ as the non-compact internal R-symmetry group of the theory.

Under this branching, the vector representation $\r{10}$ decomposes
in the obvious way as 
\be
\r{10}\ra (\r{d},\r{1})+(\r{1},(\r{10-d}))\;\;.
\ee 
The corresponding branchings of the $\r{16}$ of $SO(9,1)$, the
Majorana-Weyl spinor, can e.g.\ be read off from the table given by
Seiberg in \cite{ns16} (by reversing one of the columns). 
Subscripts $r$ and $p$ indicate real and pseudo-real
representations. 
\be
\begin{array}{c|ccc}
   & SO(9,1) &\ra & SO(d) \times SO(9-d,1)\\ \hline
d=9& \r{16}_{r} &\ra& \r{16}_{r}\\
d=8& \r{16}_{r} &\ra& (\r{8}_{s,r},\r{+1}_{r})+ (\r{8}_{c,r},\r{-1}_{r})\\
d=7& \r{16}_{r} &\ra& (\r{8}_{r},\r{2}_{r})\\
d=6& \r{16}_{r} &\ra& (\r{4},\r{2})+(\rb{4},\rb{2})\\
d=5& \r{16}_{r} &\ra& (\r4_{p},\r4_{p})\\
d=4& \r{16}_{r} &\ra& (\r2_{p},\r4_{p})+(\r2_{p}',\r4_{p}')\\
d=3& \r{16}_{r} &\ra& (\r2_{p},\r8_p)\\
d=2& \r{16}_{r} &\ra& (\r{+1},\r8_{s})+(\r{-1}=\rb{+1},\r8_{c}=
\overline{\r8_{s}})
\end{array}
\label{tp2}
\ee
Note that all the representations appearing on the right-hand side
manifestly have real structures. 
For $d=7,8,9$ the representations are simply real
and the spinors Majorana (-Weyl), for $d=3,4,5$ one has 
symplectic Majorana (-Weyl) spinors, and for $d=2,6$ the representaions
are of the form of a complex representation plus its complex conjugate
and thus also have an obvious real structure.

\subsection{Twisted $N\!=\!4$ Theories}

Among the SYM theories with 16 supercharges, the $d\!=\!4$ $N\!=\!4$ theory
is of particular interest, in particular because of its suspected
S-duality and its relation to toroidal compactifications of 
heterotic strings. Twisted
$N\!=\!4$ theories were first investigated in \cite{yamron} and then
subsequently in \cite{vw,marcus,btn2,ll}. It was found that in 
general there are three inequivalent topological twists of 
the $N\!=\!4$ theory which we will refer to as the half-twisted model
\cite{yamron,ll}, the A-model \cite{vw,btmq,dm,btn2,ll} and the
B-model \cite{marcus,btn2,ll}. 

As reviewed e.g.\ in \cite{ll}, all these investigations were
based on a (from our point of view) hybrid approach in which one
uses the compact R-symmetry group $SO(6)\sim SU(4)$ of the
Minkowskian theory to twist the Lorentz gorup $SO(4)\sim SO(3)_{L}
\times SO(3)_{R}$ of the Euclidean theory. One is thus (implicitly)
complexifying the theory and reality conditions on the fields 
have to be reimposed by hand at the end. From this point of view, 
the three $N\!=\!4$ twists are most conveniently described by the branchings
\bea
SO(6) &\ra& SO(4) \times SO(2) \sim SO(3)_{1} \times SO(3)_{2} \times SO(2)   
\non
SO(6) &\ra& SO(3)_{A}\times SO(3)_{B}    \;\;.\label{so6br}
\eea
For the half-twisted model one twists $SO(3)_{L}$ by $SO(3)_{1}$ (i.e.\
replaces $SO(3)_{L}$ by $\diag(SO(3)_{L}\times SO(3)_{1})$, for the
B-model one furthermore twists $SO(3)_{R}$ by $SO(3)_{2}$ 
and for the A-model one twists $SO(3)_{L}$ by, say, $SO(3)_{A}$. 

As we will now show
(and as might have been expected) we can completely side-step this
problem of compelexification and reality conditions by working 
directly with the hermitian Euclidean SYM action and its non-compact 
R-symmetry group $SO(5,1)$. The field content is
\be
\begin{array}{cl}
d=4 & \r{10}\ra (\r2,\r2;\r1)+(\r1,\r1;\r6)\\
    & \r{16}\ra (\r2,\r1;\r4)+(\r1,\r2;\rb4)
\end{array}
\ee
The two relevant branchings of the R-symmetry group $SO(5,1)$ are
the regular branching 
$SO(5,1)\ra SO(3)_{1}\times SO(3)_{2}\times SO(1,1)$,
\be
\begin{array}{ccl}
\r4& \ra& (\r2,\r1)^{+1}+(\r1,\r2)^{-1}\\
\rb4 & \ra & (\r2,\r1)^{-1}+(\r1,\r2)^{+1}\non
\r6 & \ra & (\r1,\r1)^{2}+(\r1,\r1)^{-2}+(\r2,\r2)^{0}
\end{array}
\ee
and the branching $SO(5,1)\ra SO(3)\times SO(2,1)$,
\be
\begin{array}{ccl}
\r4 &\ra &(\r2,\r2)\\
\rb4& \ra& (\r2,\r2)\\
\r6&\ra&(\r1,\r3)+(\r3,\r1)
\end{array}
\ee
The different twists can now be obtained as follows. Using the first 
branching and twisting $SO(3)_{L}$ by $SO(3)_{1}$ one obtains the
half-twisted model with $N_{T}\!=\!1$ (meaning one topological or scalar 
supercharge), residual global symmetry group 
$SO(3)_{2}\times SO(1,1)$, and with field content
\be
\begin{array}{rcl}
\mbox{L1-Twist}\Ra 
&\r{10}&\ra (\r2,\r2;\r1)^{0}+(\r1,\r1;\r1)^{+2,-2} +(\r2,\r1;\r2)^{0}\\
&\r{16}&\ra (\r1+\r3,\r1;\r1)^{+1}+(\r2,\r2;\r1)^{-1}
+(\r2,\r1;\r2)^{-1}+(\r1,\r2;\r2)^{+1}
\end{array}
\ee
Twisting also $SO(3)_{R}$ by $SO(3)_{2}$, one obtains the B-model
with $N_{T}\!=\!2$, residual global symmetry $SO(1,1)$ and field content
\be
\begin{array}{rcl}
\mbox{L1-R2-Twist}\Ra 
&\r{10}&\ra (\r2,\r2)^{0}+(\r1,\r1)^{+2,-2} +(\r2,\r2)^{0}\\
&\r{16}&\ra 2\times (\r1,\r1)^{+1}+(\r3,\r1)^{+1}+(\r1,\r3)^{+1}
+ 2\times (\r2,\r2)^{-1}
\end{array}
\ee
To obtain the A-model one uses the second branching
to twist the $SO(3)_{L}$ by $SO(3)$. Then $SO(2,1)$
remains as a global symmetry group and the field content is
\be
\begin{array}{rcl}
\mbox{A-Twist}\Ra 
&\r{10}&\ra (\r2,\r2;\r1)+(\r1,\r1;\r3) +(\r3,\r1;\r1)\\
&\r{16}&\ra (\r1,\r1;\r2)+(\r3,\r1;\r2)+(\r2,\r2;\r2)
\end{array}
\ee
Note that the fermionic field content is simply an $SL(2,\RR)$ doublet
of that of Donaldson theory. In particular, the reality properties
of the fields inherited from the hermitian Euclidean action are
$SL(2,\RR)$-invariant. This explains why in \cite{yamron,ll} it
was found necessary to break the (hybrid) R-symmetry group $SO(3)_{B}$, 
obtained from the branching (\ref{so6br}), down to $SO(2)$ to obtain a 
real action. 

\section{Euclidean SYM Theories on Special Holonomy Manifolds}

In this section, we will deal with Euclidean SYM theories on $d>4$
manifolds.
At present the field theoretic status of SYM theories in $d>4$ is not
entirely clear (for a recent discussion see \cite{ns16}). However, it
appears to be reasonable to believe that whenever the theory admits
a scalar (i.e.\ singlet) supercharge its associated topological field 
theory, defined
by the BRST cohomology of this supercharge, is meaningful and can be
studied in its own right. While this certainly requires further
justification, we will proceed by optimistically adopting this as a 
working hypothesis. 

Once again, for concreteness and because we are primarily interested in
$d>4$, we will focus on the SYM theories with 16 supercharges obtained
by dimensional reduction of the $N\!=\!1$ theory in $d\!=\!(9+1)$.

\subsection{General Aspects of Euclidean SYM Theories with Scalar
Supercharges}

As is well known (and we have recalled in section 2), one way to produce
a SYM theory with scalar supercharges is to twist, i.e.\ to use the 
R-symmetry group to find a new embedding of the space (-time) rotation
group into the global symmetry group, in such a way that at least one of
the supercharges of the untwisted theory becomes a singlet with respect
to the new rotation group. Now generically the rotation group (structure
group of the tangent bundle) of a $d$-manifold is $SO(d)$ while the 
R-symmetry group is $SO(9-d,1)$. Thus for $d>4$ the R-symmetry group is
simply not large enough to permit a twisting.\footnote{With the possible
exception of $d\!=\!5$ if one closes one's eyes to the difference between
$SO(5)$ and $SO(4,1)$.}
One is thus led to consider manifolds with reduced holonomy (structure)
groups $G_{M}\ss SO(d)$. For a recent survey of what is known about
holonomy groups, see e.g.\ \cite{bryant}.

Essentially the only 
interesting possibilities (of potentially `twistable' holonomy groups 
not admitting covariantly constant spinors) are K\"ahler $n$-folds with
holonomy group $G_{M}=U(n)\ss SO(2n)$. The underlying reason for why
twisting works in this case is that on a K\"ahler $n$-fold (spin$_{c}$)
spinors can be identified with twisted differential forms (more precisely
one has
\be
(\Bbb{S}^{+}\oplus\Bbb{S}^{-})\otimes K_{M}^{+1/2}\simeq \Omega_{M}^{*,0}
\ee
where $K_{M}$ is the canonical bundle of $M$ and $\Bbb{S}^{\pm}$ are the
spin-bundles) and that one can use $U(1)$ subgroups of the R-symmetry group 
to untwist them, i.e.\ to mimic tensoring by square roots of $K_{M}$
(see also \cite{lmns}).
We will explain this in somewhat more detail below.

Another way to produce a SYM theory with scalar supercharges is to place 
it on a spin-manifold $M$ admitting covariantly constant spinors. This
means that one is once again led to consider reduced holonomy groups
$G_{M}\ss SO(d)$, this time such that one obtains at least one 
$G_{M}$-singlet
from the branching of the spinor representation under $SO(d)\ra G_{M}$.
This is of course a well known property of Calabi-Yau manifolds
(and the principal reason for why they appear in the context of
string compactifications in the first place). 
In $d<10$, the only other non-trivial
possibilities are known to be K3-surfaces (i.e.\ Calabi-Yau or
hyper-K\"ahler 2-folds with holonomy $SU(2)$), hyper-K\"ahler 4-folds 
with holonomy group $Sp(2)$, and the exceptional $G_{2}$ and 
$Spin(7)$ holonomy Joyce seven- and eight-manifolds. 

Typically,
on such a manifold the spinor bundle becomes isomorphic to a 
tensor bundle, as exemplified by the branching
\be
SO(7)\ra G_{2}\;:\;\;\;\;\;\;\r8\ra\r1+\r7\;\;,\label{sog2}
\ee
which means that a spinor on a $G_{2}$-manifold can be regarded as a
pair $(\eta,\psi_{\mu})$ consisting of a scalar and a vector field.
Thus a SYM theory on such a manifold acquires all (or at least many of)
the characteristics of a cohomological topological field theory all by
itself, i.e.\ without twisting it explicitly. Some aspects of these
theories (notably for $G_{M}=SU(4)$ and $G_{M}=Spin(7)$) have been studied
in \cite{bks} (see also \cite{aos}), where the relation to the 
generalized self-duality conditions of \cite{cdfn,ward} and the moduli 
problems studied in \cite{dt} were pointed out. We will survey the
possibilities that arise in this way below in the light of 
the considerations of section 2. 

Among the admissible metric holonomy groups there are also 
pseudo-versions of the groups above which can occur as reductions
of pseudo-Riemannian manifolds \cite{bryant}, such as 
$U(p,q)\ss SO(2p,2q)$ and a non-compact counterpart of the 
reduction (\ref{sog2}), namely $G_{2(1)}\ss SO(4,3)$. If interest
in exotic signatures persists in the physics literature, then 
perhaps these manifolds will also come to play a role.

\subsection{SYM Theories on K\"ahler and Calabi-Yau Manifolds}

In order to determine the spectrum (and possible twists) of SYM
theories on K\"ahler and Calabi-Yau $n$-folds, we require the
branching of the vector and spinor representations under 
$SO(2n)\ra U(n)$. 
As on a K\"ahler manifold a one-form can be decomposed as the sum of a
$(1,0)$ and a $(0,1)$ form, the general structure for the decomposition
of the fundamental representation $\r{2n}$ of $SO(2n)$ under the 
branching $SO(2n)\ra U(n)$ is
\be
\r{2n}\ra \r{n}^{+q}+\rb{n}^{-q}\;\;.
\ee
Here $\r{n}$ and $\rb{n}$ designate representations of $SU(n)$ and $q$
is a $U(1)$-charge. It is conventionally normalized to be the smallest
integers such that all representations have integer $U(1)$-charge. For
$n=2,3,4$ one has $q=1,2,1$ respectively. Thus, as  $\r{n}^{+q}$ 
corresponds to $(1,0)$-forms, the canonical line bundle $K$ and its square 
root $K^{1/2}$ then correspond to the representations $\r1^{+nq}$ and 
$\r1^{+nq/2}$ respectively.

To simplify the notation,
occasionally when one $SU(n)$ representations $\r{d}$ appears with 
several $U(1)$ charges $a,b,c,\ldots$, we will group them together and
write them as $\r{d}^{a,b,c,\ldots}$. With these notations, one then
has the following table for the relevant (vector, spinor, two-form)
representations:
\be
\begin{array}{cccc}
SO(4)\ra U(2) & \r4=(\r2,\r2) & \ra & \r2^{+1} + \r2^{-1} \\
              & \r2_{L} = (\r2,\r1) & \ra & \r2^{0}\\
              & \r2_{R} = (\r1,\r2) & \ra & \r1^{+1} + \r1^{-1}\\
              &\r6 = (\r3,\r1)+(\r1,\r3) & \ra & \r3^{0}+\r1^{+2,0,-2}\\
SO(6)\ra U(3) &\r6 &\ra & \r3^{2} + \rb3^{-2}\\
              &\r4 &\ra & \r1^{+3} + \r3^{-1}\\
              &\rb4 &\ra & \r1^{-3} + \rb3^{+1}\\
              & \r{15} & \ra & \r1^{0} + \r3^{-4} +\rb3^{+4} +\r8^{0}\\
SO(8)\ra U(4) & \r8_{v} & \ra &\r4^{+1} +\rb{4}^{-1}\\
              & \r8_{s} & \ra & \r1^{+2,-2} +\r6^{0}\\
              & \r8_{c} & \ra &\r4^{-1} +\rb{4}^{+1}\\
              & \r{28} & \ra &\r1^{0} + \r6^{+2,-2} + \r{15}^{0}
\end{array} 
\label{t1}
\ee
Here for $SO(4)$ we have assumed that the reduction to $U(2)$ proceeds
via $SU(2)_{L}\times SU(2)_{R} \ra SU(2)_{L}\times U(1)_{R}$. Note that
in each case, the decomposition of the two-forms contains a singlet
$\r1^{0}$. This corresponds to the covariantly constant K\"ahler
two-form.  The other terms in the two-form decomposition correspond to the
$(2,0)$ and $(0,2)$ forms and to the $(1,1)$-forms not proportional 
to the K\"ahler form. The latter transform in the adjoint of $SU(n)$, 
as follows from $\r{n}\times\rb{n} = \r1 + (\r{n^{2}-1})$.  
Upon restriction to Calabi-Yau $n$-folds with holonomy group $SU(n)$,
one finds two covariantly constant spinors of the same chirality
for $n=2,4$, and two of opposite chirality for $n=3$. 

It follows from the above that on a K\"ahler $n$-fold for $n=2,3,4$ one has
the following relation between differential forms and twisted spinors
(spin$_{c}$ spinors - these exist on any K\"ahler manifold $M$ even when
$M$ is not spin):
\be
\begin{array}{cll}
n=2 & \Bbb{S}^{+}\otimes K^{+1/2} \simeq \r1^{0} +\r1^{+2} &=\Omega^{even,0}\\
    & \Bbb{S}^{+}\otimes K^{-1/2} \simeq \r1^{0} +\r1^{-2} &=\Omega^{0,even}\\
    & \Bbb{S}^{-}\otimes K^{+1/2} \simeq \r2^{+1} &=\Omega^{odd,0}\\
    & \Bbb{S}^{-}\otimes K^{-1/2} \simeq \r2^{-1} &=\Omega^{0,odd}\\
n=3 & \Bbb{S}^{+}\otimes K^{+1/2} \simeq \r1^{+6}+\r3^{+2} &=\Omega^{odd,0}\\
    & \Bbb{S}^{+}\otimes K^{-1/2} \simeq \r1^{0}+\r3^{-4} &=\Omega^{0,even}\\
    & \Bbb{S}^{-}\otimes K^{+1/2} \simeq \r1^{0} +\rb3^{+4} &=\Omega^{even,0}\\
    & \Bbb{S}^{-}\otimes K^{-1/2} \simeq \r1^{-6}+\rb3^{-2} &=\Omega^{0,odd}\\
n=4 & \Bbb{S}^{+}\otimes K^{+1/2} \simeq \r1^{+4}+\r1^{0} +\r6^{2} 
&=\Omega^{even,0}\\
    & \Bbb{S}^{+}\otimes K^{-1/2} \simeq \r1^{0} + \r1^{-4} +\r6^{-2}
&=\Omega^{0,even}\\
    & \Bbb{S}^{-}\otimes K^{+1/2} \simeq \r4^{+1}+\rb4^{+3} &=\Omega^{odd,0}\\
    & \Bbb{S}^{-}\otimes K^{-1/2} \simeq \r4^{-3}+\rb4^{-1} &=\Omega^{0,odd}
\end{array}
\label{t2}
\ee
On a Calabi-Yau manifold, $K$ is trivial, and choosing the trivial
square root of $K$, the above twist is trivial. Thus on a Calabi-Yau
manifold, one can identify spinors directly with differntial forms.

The above correspondence, valid at the level of complex
representations, does not yet take into account the effect of being able
to impose reality (Majorana) conditions on the spinors in certain cases
(notably for $n=4$). These reality properties can be read off from 
table (\ref{tp2}), and combining this with (\ref{t1}) one finds that the 
$U(n)\times SO(9-2n,1)$ field content of the $N\!=\!1$ $d\!=\!(9+1)$ 
SYM theory reduced to a K\"ahler $n$-fold is
\be
\begin{array}{cl}
n=2 & \r{10}\ra (\r2^{+1}+\r2^{-1};\r1)+(\r1^{0};\r6)\\
    & \r{16}\ra (\r2^{0};\r4)+(\r1^{+}+\r1^{-1};\r4')\\
n=3 & \r{10}\ra (\r3^{+2}+\rb3^{-2};\r1)+(\r1^{0};\r4)\\
    & \r{16}\ra (\r1^{+3}+\r3^{-1};\r2)+(\r1^{-3}+\rb3^{+1};\rb2)\\
n=4 & \r{10}\ra \r4^{+1;0}+\rb4^{-1;0}+\r1^{0;+2}+\r1^{0;-2} \\
    & \r{16}\ra \r4^{-1;+1}+\rb4^{+1;+1}+\r6^{0;-1}+\r1^{+2;-1}+\r1^{-2;-1}
\end{array}
\label{t4}
\ee
In a number of respects, the case $n=2$ is somewhat special. First of all,
one does not need a reduced holonomy group in order to be able to twist
the theory. Secondly, these three standard twists are well understood and
can readily be specialized to K\"ahler manifolds if this is required (see
e.g.\ \cite{vw}). And finally, as is also well known, on a hyper-K\"ahler
(K3) 2-fold, one does not need to twist at all and one obtains directly
an $N_{T}\!=\!8$ theory. This is consistent with the fact that a K3 is known 
to preserve 1/2 of the sueprsymemtries, thus reducing the 16 supercharges 
of the flat space theory to 8. 

For $n=3$, the R-symmetry group is $SO(3,1)\sim SL(2,\CC)$ 
and in this case the theory can be twisted via a suitable 
$U(1)$-subgroup of $SL(2,\CC)$ (chosen such as to cancel 
the $U(1)$-charges of the $SU(3)$ singlets appearing in 
the spinor branching). Concretely,
with a convenient normalization of the $U(1)$ charges,
one has
\be
\begin{array}{ccccl}
\r{16} & \ra & (\r1^{+3}+\r3^{-1},\r2)+(\r1^{-3}+\rb3^{+1},\rb2) 
       & \mbox{of} & U(3) \times SL(2,\CC)\\
    & \ra & (\r1^{+3}+\r3^{-1},\r1^{\pm 3})+(\r1^{-3}+\rb3^{+1},\r1^{\pm 3}) 
       & \mbox{of} & U(3) \times U(1)\\
    & \ra & \r1^{+6,0,0,-6}+\r3^{+2}+\r3^{-4}+\rb3^{+4}+ \rb3^{-2}  & 
\mbox{of} & U(3)\;\; \mbox{by twisting} \\
    & \simeq& \Omega^{*,0}\oplus \Omega^{0,*} &&
\end{array}
\ee
Thus this theory has $N_{T}\!=\!2$ and 
a residual global $\CC^{*}$ symmetry, the centralizer of $U(1)\ss SL(2,\CC)$.
The reality conditions on the spnors imply the obvious reality
condition $(\omega^{p,0})^{*}=\omega^{0,p}$.

If one introduces the complex Grassmann odd fields $\eta^{0,0},
\psi^{1,0},\chi^{2,0},\eta^{3,0}$ and their complex conjugates,
then the fermionic action can schematically be written as
\be
S_{F}=\int \eta^{0,3}\del_{A}\chi^{2,0} 
           +\omega^{2}\psi^{0,1}\del_{A}^{*}\chi^{2,0}
           +\omega^{3}\eta^{0,0}\del_{A}^{*}\psi^{1,0} + c.c.
\ee
Upon twisting, the bosonic spectrum consists of the gauge field,
two scalars, a (3,0) and a (0,3) form, and the action provides
a (topological) field theory description of the DUY (Donaldson, 
Uhlenbeck, Yau) equations
\bea
F^{2,0}=F^{0,2}&=&0\non
\omega^{n-1}\wedge F^{1,1}&=&0
\eea
for $n=3$, describing stable vector bundles. Its one-loop
approximation can be described by the generalized bc-systems of 
\cite{lmns} and is thus governed by the Ray-Singer torsion of $M$.
If the 3-fold is Calabi-Yau,
then once again no twisting is necessary, the theory exhibits an
$N_{T}\!=\!4=16/4$ topological symmetry (and the full $SL(2,\CC)$ R-symmetry)
and coincides with the theory studied in \cite{bks}. 

Finally, for $n=4$, the R-symmetry group is $SO(1,1)$ and hence cannot
be used directly (i.e.\ while preserving the reality conditions)
to twist the $U(1)$-subgroup of $U(4)$. On a Calabi-Yau 4-fold, however,
one finds the $H$-theory of \cite{bks}, an $N_{T}\!=\!2$ theory describing
a holomorphic analogue \cite{dt} of Donaldson theory, i.e.\ gauge fields
satisfying the holomorphic self-duality condition 
\be
\star F^{0,2}= F^{0,2}\;\;,\label{holins}
\ee
where $\star: \Omega^{0,2}\ra \Omega^{0,2}$
satifying $\star^{2}=1$ is defined via the holomorphic 4-form. This
self-duality condition arises as a consequence of the reality property
of the $\r6=(\r4\wedge\r4)$ appearing in the branching of the $\r8_{s}$.

\subsection{A Brief Survey of SYM Theories on Other Special Holonomy 
Manifolds}

Apart from K\"ahler and Calabi-Yau manifolds, discussed above, the only
other manifolds on which SYM theories acquire scalar supercharges 
are $Spin(7)$ (or octonionic) and hyper-K\"ahler eight-manifolds and
$G_{2}$ seven-manifolds. The first case has been discussed at length 
in \cite{bks,aos} to which we refer for details. Here the moduli space
in question is defined by the equation \cite{cdfn}
\be
F_{8i}=c_{ijk}F_{jk}\label{octins}
\ee
where the $c_{ijk}$ are the octonionic structure constants. This equation
can be understood more invariantly from the branching $\r8_{s}\ra\r7+\r1$
which exhibits the one covariantly constant spinor $\zeta$
($\r8_{v}$ and $\r8_{c}$ remain irreducible under $SO(8)\ra SO(7)$). 
Consequently the two-form decomposition is $\r{28}\ra\r{21}+\r7$,
corresponding to the two eigenspaces of the operator 
\be
T^{ijkl}=\zeta^{T}\ga^{[i}\ga^{j}\ga^{k}\ga^{l]}\zeta
\ee
on two-forms, and the seven equations (\ref{octins}) represent a projection
onto the $\r{21}$. As anticipated in \cite{bks,aos}, the field theory 
associated to this moduli problem arises from the eight-dimensional 
Euclidean SYM theory, 

The $G_{2}$ case arises from a straightforward dimensional reduction 
of the octonionic theory: the branching (\ref{sog2}) induces the 
two-form decomposition $\r{21}\ra\r{14}+\r7$ and the reduction of 
(\ref{octins}) projects onto the $\r{14}$ \cite{cdfn}. 

Finally, we study the hyper-K\"ahler case in a little more 
detail. Let us initially consider the branching 
$SO(8)\ra Sp(2)\times Sp(1)$ (so this would correspond to a quaternionic 
K\"ahler manifold).  The vector, spinor, and two-form representations
branch to
\bea
\r8_{v} &\ra& (\r4,\r2)\non
\r8_{s} &\ra& (\r5,\r1) + (\r1,\r3)\non
\r8_{c} &\ra& (\r4,\r2)\non
\r{28}  &\ra& (\r1,\r3)+(\r{10},\r1) + (\r5,\r3)
\eea
Thus a hyper-K\"ahler eight-manifold admits three covariantly
constant spinors (of the same chirality).
The defining representation of $Sp(n)$ on $\HH^{n}$ can be regarded
as a representation $\r{2n}$ on $\CC^{2n}$. The holonomy group of an
eight-manifold being reduced to $Sp(2)\times Sp(1)$ means that the 
(co-)vector $V_{\mu}$ can be written as $V_{aA}$ where $a=1,2,3,4$ 
and $A=1,2$. In other words, $\r8_{v}\ra(\r4,\r2)$.
The reality condition $V_{\mu}^{*}=V_{\mu}$ can be written
as
\be
V^{*,aA}\equiv V_{aA}^{*} = \epsilon^{ab}\epsilon^{AB}V_{bB}
\ee
($\r{4}$ and $\r{2}$ being pseudo-real, $(\r4,\r2)$ has a real structure).
As $\r8_{c}$ branches in the same way as $\r8_{v}$, this also explains
the real structure induced by $\r8_{c}$. 

The $\r8_{s}$, on the other hand, branches to $(\r5,\r1)+(\r1,\r3)$.
The second term corresponds to a real
symmetric tensor $\eta_{(AB)}$ in the $\r2\symx\r2 = \r3$ of $Sp(1)=SU(2)$.
The $\r{5}$ of $Sp(2)$ has the following 
interpretation. The $\wedge^{2}\r4$ of $Sp(2)$ is a six-dimensional 
representation which is, however, reducible as $Sp(2)$ leaves 
invariant the symplectic form $\epsilon_{ab}$. Thus one has the
decomposition 
\bea
\r4\wedge\r4 &=& \r1 + \r5\non
\omega_{ab} &=& \epsilon_{ab}\xi + \chi_{ab}\;\;,
\eea
where $\chi_{ab}$ is anti-symmetric and trace free, 
$\epsilon^{ab}\chi_{ab}= 0$.
The reality condition on $8_{s}$ thus imposes the `self-duality'
condition
\be
\chi_{ab}^{*}=\epsilon^{ac}{\epsilon}^{bd}\chi_{cd}\;\;.
\ee
on $\chi_{ab}$ (the hermitian metric is implicit in this equation).

The relation between
the three covariantly constant spinors $\eta_{r}$, $r=1,2,3$,
normalized such that $\eta_{r}^{T}\eta_{r}=1$, and the three covariantly
constant K\"ahler forms on a hyper-K\"ahler manifold is
\be
      \Omega^{r}_{ij} =\epsilon^{rst}\eta_{s}^{T}\gamma_{ij}\eta_{t}/2
\ee
where the $\gamma_{i}$ are $SO(8)$ gamma matrices. This is also reflected in
the branching of the $\r{28}$ exhibited above, which contains three singlets
from the $(\r1,\r3)$ when the holonomy is $Sp(2)$.

A general two-form on a quaternionic K\"ahler manifold, say the curvature
of a connection $A_{aA}$, has the form $F_{aAbB}$. 
Decomposing this into its irreducible pieces according to
\bea
\r{28}&=&\r8_{v}\wedge\r8_{v} \ra  (\r4,\r2)\wedge(\r4,\r2) \non
      &=& (\r4\wedge\r4,\r2\symx\r2) + (\r4\symx\r4,\r2\wedge\r2)\non
      &=& (\r1,\r3) + (\r5,\r3) + (\r{10},\r1)\;\;,
\eea
one obtains (in the notation of \cite{ward})
\be
F_{aAbB} = \epsilon_{ab}G_{AB}+K_{abAB}+ \epsilon_{AB}H_{ab}\;\;,
\ee
where $G$ and $H$ are symmetric and 
\be
K_{abAB}=\trac{1}{2}(F_{aAbB}+F_{aBbA}) - \epsilon_{ab} G_{AB}
\ee
is symmetric in $(AB)$ and skew trace-free in $(ab)$.

In \cite{ward}, Ward discusses $Sp(2)\times Sp(1)$ invariant integrable
equations for gauge fields, and he obtains the set of 18 equations
\be
G_{AB}=K_{abAB}=0
\ee
as the integrability conditions $\pi^{A}\pi^{B}F_{aAbB}=0$
for the equations $\pi^{A}D_{aA}\psi=0$. On a hyper-K\"ahler manifold,
one can split these equations invariantly into three parts and consider one
of them seperately, say
\be
G_{12}+G_{21} =K_{ab12}+K_{ab21} =0\;\;.\label{hksd}
\ee
It is this set of $(5+1)$ equations that one obtains
from SYM theory. Indeed, the field content is (displaying
also the $SO(1,1)$ R-symmetry quantum numbers)
\be
\begin{array}{ccl}
\r{10} &\ra& 2\times \r4^{0} + \r1^{+2,-2}\\
\r{16} &\ra& \r5^{-1}+3\times\r1^{-1}+2\times\r4^{+1}
\end{array}
\ee
Here we recognise, first of all, the gauge field $A_{aA}=(A_{a1},A_{a2})$
and its ghost-number one superpartner $\psi_{aA}$. Secondly, one sees
the two scalars, say $\eta_{11}$ and $\eta_{22}$ required to gauge fix
the $\psi_{aA}$. And thirdly there are the anti-ghosts $\eta$ and 
$\chi_{ab}$ imposing the self-duality conditions (\ref{hksd}). 
Finally, as in Donaldson theory there are a bosonic anti-ghost and a 
ghost-for-ghost. The theory has has the unusal property of possessing 
an $N_{T}\!=\!3$ symmetry, all supercharges carrying the same $SO(1,1)$
ghost number and tranforming as an $Sp(1)$ triplet.

These equations are the reduction under $SU(4)\ra Sp(2)$ of the
holomorphic self-duality equations of the $SU(4)$ H-theory \cite{bks}
discussed above (by an appropriate choice of complex structure). 
As such, thinking of the latter as a holomoprhic
version of Donaldson theory, the hyper-K\"ahler equations may play 
a role akin to that of instanton equations on symplectic or K\"ahler 
four-manifolds.

\appendix

\section{Notation}

Our conventions for $\ga$-matrices in $d\!=\!4$ and $d\!=\!(5+1)$ are the
following: For $d\!=\!4$ the gamma matrices are taken to be hermitian 
and in terms of Pauli matrices $\sigma_{a}$ we choose
\bea
\ga_{a} &=& \sigma_{1}\otimes \sigma_{a}\;\;\;\;\;\;a=1,2,3\non
\ga_{4} &=& \sigma_{2}\otimes \II\;\;,
\eea
so that $\ga_{5}$ is diagonal,
\be
\ga_{5} = \sigma_{3}\otimes \II\;\;.
\ee
The `charge conjugation' matrix $C = \ga_{2}\ga_{4}$
satisfies $C^{T}= -C$ and
\be
C\ga_{m} = \ga_{m}^{T}C
\ee
where $m=1, \dots, 5$.

The conjugate of a spinor is
\be
\overline{\Psi} = \Psi^{*T} ,
\ee
while the chiral decomposition is
\be
\Psi = \left( \begin{array}{c}
\psi_{+} \\
\psi_{-}
\end{array}\right)
\ee

For $d\!=\!(5+1)$ Minkowski space we choose
\bea
\Gamma^{m}&=&\sigma_{1}\otimes \ga_{m}\;\;\;\;\;\;m=1,\ldots,5\non
\Gamma^{0}&=&i\sigma_{2}\otimes \II_{4}\;\;,
\eea
so that $\Gamma^{7}$ is also diagonal,
\be
\Gamma^{7}=\sigma_{3}\otimes \II_{4}\;\;. 
\ee
Lorentz generators are denoted by
\be
\Sigma^{MN} = \trac{1}{4}[\Gamma^{M},\Gamma^{N}]\;\;.
\ee
The charge conjugation matrix ${\cal C}$ satisfying
\be
{\cal C}\Gamma_{m} = \Gamma_{m}^{T}{\cal C}, \;\;\;
{\cal C} \Gamma_{0} = - \Gamma_{0}^{T}{\cal C} 
\ee
as well as
\be
{\cal C}^{T} = -{\cal C} = {\cal C}^{-1}
\ee
is
\be
{\cal C} = \II_{2}\otimes C\;\;.
\ee

\rnc{\Large}{\normalsize}


\begin{thebibliography}{00}
\addcontentsline{toc}{section}{References}
\frenchspacing
\small
\addtolength{\itemsep}{-4pt}
\bibitem{zumino} B. Zumino, {\em Euclidean supersymmetry and the
many-instanton problem}, Phys. Lett. 69B, (1977) 369-371.
\bibitem{vanN} P. van Nieuwenhuizen, A. Waldron {\em On Euclidean spinors 
and Wick rotations}, Phys. Lett. B389 (1996) 29-36, {\tt hep-th/9608174};
{\em A continuous Wick rotation for spinor fields and supersymemtry in
Euclidean space}, {\tt hep-th/9611043}.
\bibitem{ewdon} E. Witten, {\em Topological quantum field theory},
Commun. Math. Phys. 117 (1988) 353.   
\bibitem{yamron} J. Yamron, {\em Topological actions from twisted
supersymmetric theories}, Phys. Lett. B213 (1988) 325-330.
\bibitem{vw} C. Vafa, E. Witten, {\em A strong coupling test of S-duality}, 
Nucl. Phys. B431 (1994) 3-77, {\tt hep-th/9408074}.
\bibitem{marcus} N. Marcus, {\em The other topological twisting of $N\!=\!4$
Yang-Mills}, Nucl. Phys. B452 (1995) 331, {\tt hep-th/9506002}.
\bibitem{bsv} M. Bershadsky, V. Sadov, C. Vafa, {\em D-branes and topological
field theories}, Nucl. Phys. B463 (1996) 420, {\tt hep-th/9511222}.
\bibitem{btn2} M. Blau, G. Thompson, {\em Aspects of $N_{T}\geq 2$
topological gauge theories and D-branes}, Nucl. Phys. B492 (1997) 545-590,
{\tt hep-th/9612143}.
\bibitem{ll} J.M.F. Labastida, C. Lozano, {\em Mathai-Quillen formulation
of twisted $N\!=\!4$ supersymmetric gauge theories in four dimensions}, 
{\tt hep-th/9702106}.
\bibitem{ewpb} E. Witten, {\em Bound states of strings and $p$-branes},
Nucl. Phys. B460 (1996) 335. {\tt hep-th/9510135}.
\bibitem{bks} L Baulieu, H. Kanno, I.M. Singer, {\em Special quantum field
theories in eight and other dimensions}, {\tt hep-th/9704167}; {\em
Cohomological Yang-Mills theory in eight dimensions}, {\tt hep-th/9705127}.
\bibitem{aos} B.S. Acharya, M. O'Loughlin, B. Spence, {\em Higher 
diemnsional analogues of Donaldson-Witten theory}, {\tt hep-th/9705138}.
\bibitem{aos2}
B.S. Acharya, M. O'Loughlin, {\em Self-duality in $D\leq 8$-dimensional
Euclidean gravity}, {\tt hep-th/9612182}.
\bibitem{cdfn} E. Corrigan, C. Devchand, D. Fairlie, J. Nuyts, 
{\em First-order equations for gauge fields in spaces of dimension 
greater than four}, Nucl. Phys. B214 (1983) 452-464.  
\bibitem{ward} R.S. Ward, {\em
Completely solvable  gauge-field equations in dimensions greater than four},
 Nucl. Phys. B236 (1984) 381-396. 
\bibitem{dt} S.K. Donaldson, R.P. Thomas, {\em Gauge theory in higher
dimensions}, Oxford preprint
\bibitem{kt} T. Kugo, P. Townsend, {\em Supersymmetry and the division 
algebras}, Nucl. Phys. B221 (1983) 357-380.
\bibitem{bjsv} M. Bershadsky, A. Johansen, V. Sadov, C. Vafa, 
{\em Topological reduction of 4d SYM to 2d $\sigma$-models}, Nucl. Phys.
B448 (1995) 166, {\tt hep-th/9501096}.
\bibitem{ns16} N. Seiberg, {\em Notes on theories with 16 supercharges},
{\tt hep-th/9705117}.
\bibitem{btmq} M. Blau, G. Thompson,  {\em $N\!=\!2$ topological gauge theory,
the Euler characteristic of moduli spaces, and the Casson invariant},
Commun. Math. Phys. 152 (1993) 41-71, {\tt hep-th/9112012}.
\bibitem{dm} R. Dijkgraaf, G. Moore, {\em Balanced topological field
theories}, Commun. Math. Phys. 185 (1997) 411-440, {\tt hep-th/9608169}.
\bibitem{bryant}  R. Bryant, {\em Classical, exceptional, and exotic
holonomies: a status report}, in {\em 
Actes de la Table Ronde de G\'eom\'etrie Diff\'erentielle en 
l'Honneur de Marcel Berger}, Collection SMF S\'eminaires and Congr\`es 1 
(1996) (Soc.  Math. de France), 93-166, \\available as
{\tt http://www.math.duke.edu/preprints/95-10.dvi}.
\bibitem{lmns} A. Losev, G. Moore, N. Nekrasov, S. Shatashvili,
{\em Chiral Lagrangians, anomalies, supersymmetry, and holomorphy},
Nucl. Phys. B484 (1997) 196-222, {\tt hep-th/9606082}.
\end{thebibliography}
\end{document}